# Design, fabrication and characterization of Computer Generated Holograms for anti-counterfeiting applications using OAM beams as light decoders

Gianluca Ruffato[1,2], Roberto Rossi[1,2], Michele Massari[1,2], Erfan Mafakheri[1,2], Pietro Capaldo[2,3] and Filippo Romanato[1,2,3*]

[1]Department of Physics and Astronomy 'G. Galilei', University of Padova, via Marzolo 8, 35131 Padova, Italy

[2]LaNN, Laboratory for Nanofabrication of Nanodevices, EcamRicert, Corso Stati Uniti 4, 35127 Padova, Italy.

[3]CNR-INFM TASC IOM National Laboratory, S.S. 14 Km 163.5, 34012 Basovizza, Trieste, Italy

*Corresponding author: filippo.romanato@unipd.it

ABSTRACT

In this paper, we present the design, fabrication and optical characterization of computer-generated holograms (CGH) encoding information for light beams carrying orbital angular momentum (OAM). Through the use of a numerical code, based on an iterative Fourier transform algorithm, a phase-only diffractive optical element (PH-DOE) specifically designed for OAM illumination has been computed, fabricated and tested. In order to shape the incident beam into a helicoidal phase profile and generate light carrying phase singularities, a method based on transmission through high-order spiral phase plates (SPPs) has been used. The phase pattern of the designed holographic DOEs has been fabricated using high-resolution Electron-Beam Lithography (EBL) over glass substrates coated with a positive photoresist layer (polymethylmethacrylate). To the best of our knowledge, the present study is the first attempt, in a comprehensive work, to design, fabricate and characterize computer-generated holograms encoding information for structured light carrying OAM and phase singularities. These optical devices appear promising as high-security optical elements for anti-counterfeiting applications.

INTRODUCTION

The increasing need for ID security and brand protection is driving global adoption of sophisticated technologies to provide considerable and effective barriers to counterfeit through the use of security holograms [1]. In response to this demand, holographic industries devote a big afford to enhance intrinsic material properties and fabrication complexity [2]. Likewise, fraud and counterfeiting techniques evolve and the majority of conventional security features are compromised. In order to prevent forgery attempts, the holographic patterns are constantly implemented by increasing the security level, reaching sometimes a degree of sophistication that overwhelms the original intent of fraud prevention. Even if today the development of micro-ad nano-scale fabrication techniques can produce optical and physical features virtually impossible to counterfeit, on the other hand it is important to provide easy detection methods to the examiners [3].

With this intent, in this work we focus on the degree of freedom represented by the decoding illumination and we engineer a new type of diffractive optical elements (DOEs) which correctly decode visual information only when illuminated with light owning specific spatial distributions of intensity and phase. As numerical analysis and preliminary experimental results suggested [4], it is possible to consider these high-security optical elements as a further advancement in anti-fraud technologies.

Traditionally, a hologram is referred to as a physically-recorded interference pattern between a coherent reference beam and the wave scattered by an object [5]. The presence of the actual physical object is nowadays not necessary since the mathematical relations between object and image fields can be implemented numerically. In computer-generated holograms (CGHs), in fact, the hologram pattern is numerically designed and optimized, in terms of signal to noise ratio (SNR) and diffraction efficiency ($\eta$) of the reconstructed image [6]. The physical process that allows the reconstruction of the image in far-field through a diffraction mechanism occurring in the holographic pattern is mathematically expressed by the Fresnel-Kirchhoff diffraction equation [7]. In the paraxial approximation, this relation provides the diffraction field $O(x,y,z)$ generated at a distance $z$ from a diffractive optical element, and it is given by:

$$O(x,y,z) = \frac{e^{ikz}}{i\lambda z} \iint_S U^i(x',y')G(x',y') e^{ik\frac{(x-x')^2+(y-y')^2}{2z}} dx'dy' \qquad (1)$$

where $U^i(x', y')$ is the complex field incident on the holographic plane, $G(x', y')$ is the DOE transmission function, $(x', y')$ and $(x, y)$ are the Cartesian coordinates on the holographic plane and on the image plane respectively, and $k=2\pi/\lambda$ is the incident wave vector, being $\lambda$ the incident wavelength. By developing the two square terms in the exponential contribution inside the integral in eq. (1), we get the following form:

$$O(x,y,z) = \frac{e^{ikz}}{i\lambda z} e^{ik\frac{x^2+y^2}{2z}} FT[A^*]\left(\frac{x}{\lambda z}, \frac{y}{\lambda z}\right) \qquad (2)$$

where $FT$ stands for the Fourier Transform operator. Therefore the diffraction field $O(x,y,z)$ is related to the Fourier Transform of a modified hologram transmission function $A^*$, calculated at the spatial frequencies $(x/\lambda z, y/\lambda z)$. The transmission function $A^*$ is defined as:

$$A^*(x,y) = U^i(x,y)G(x,y)\exp\left(ik\frac{x'^2+y'^2}{2z}\right) \qquad (3)$$

and results to be the product between the hologram phase function, the incident field on the hologram plane, and the Fresnel phase factor. Previous equations serve as the basis for the computation of Fresnel computer generated holograms with specific incident illumination.

Diffractive optics can be designed to work either in transmission or in reflection and can be engineered to manipulate either the phase or the amplitude (or both) of the input wave. Due to their higher efficiency, phase-only holograms are far more preferable. The phase control of the holographic pattern is expressed by its phase function $\varphi(x, y)$, which is given below for the two different configurations. For transmission holograms, we have:

$$\varphi(x,y) = \frac{2\pi}{\lambda} \cos(\vartheta_i^*)[n(\lambda)-1]d(x,y) \qquad (4)$$

being $\vartheta_i^*$ the propagation angle inside the hologram medium, $n(\lambda)$ the refractive index for the given wavelength, $d(x, y)$ the local thickness of the hologram at the position $(x, y)$. In case of holograms working in reflection, we have instead:

$$\varphi(x,y) = \frac{4\pi}{\lambda} \cos(\vartheta_i) h(x,y) \qquad (5)$$

defining $\vartheta_i$ as the incident angle of illumination and $h(x, y)$ as the depth of the hologram pattern at the coordinates $(x, y)$. A schematic representation of the light path interacting either in transmission or in reflection with a multilevel phase-only diffractive optical element (PO-DOE) is shown in Figure 1.

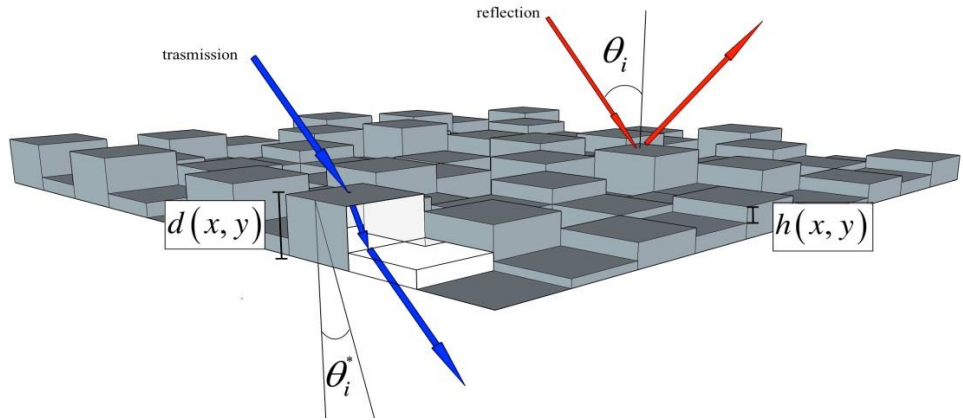

**Figure 1.** Schematic representation of the light path interacting either in transmission or in reflection with a multilevel phase-only diffractive optical element (PO-DOE).

Once the phase pattern of the hologram has been computed, last equations allow the implementation of the holographic nanostructured surface.

After the seminal paper of Allen and co-workers in 1992 [8], it is known that light beams characterized by helicoidal phase-fronts possess a well-defined orbital angular momentum (OAM). Such beams are characterized by a phase term $exp(i\ell\varphi)$, being $\ell$ the amount of OAM carried by each single photon in units of $\hbar$. Since then, light beams carrying OAM have gained increasing attention due to the wide range of uses and different possible applications [9-12], such as: particle trapping [13] and tweezing [14], phase contrast microscopy [15], STED microscopy [16], quantum-key distribution [17] and telecommunications [18].

In the paraxial regime, an OAM beam can be described in terms of Laguerre-Gaussian Modes (LG) characterized by two indices $\ell$ and $p$, the azimuthal and radial index respectively. The azimuthal index $\ell$, corresponding to the topological charge of the embedded phase singularity, represents the number of intertwined helical wave-fronts. The index $p$ represents the number of radial nodes on a plane perpendicular to the direction of propagation and it is related to the distribution of the intensity pattern in $p+1$ concentric rings around the central dark zone of the phase singularity.

Different techniques have been presented to tailor the orbital angular momentum of a beam, such as astigmatic mode converters [19], fork-holograms [20] and $q$-plates [21]. In this work, we use a method based on transmission through spiral phase plates (SPPs). SPPs are phase optical elements looking like spiral staircases, which are able to shape an incident Gaussian beam into an OAM beam, as first shown by Beijersbergen *et al.* [22]. Common SPPs are transparent optical elements whose thickness $h$ increases as a function of the azimuthal coordinate according to:

$$h(r,\varphi) = \ell \frac{\varphi}{2\pi} \frac{\lambda}{n_{SPP} - n_0} \tag{6}$$

where $n_{SPP}$ is the refractive index of the SPP material, $n_o$ is the refractive index of the surrounding medium, usually air, and $\lambda$ is the impinging wavelength.

Detailed work has been done in order to optimize both design and fabrication procedures for the generation of high-order OAM beams with non-zero radial index [23]. This is feasible by introducing radial π-discontinuities on the SPP phase pattern $\Omega_{SPP}$:

$$\Omega_{SPP}(r,\varphi) = \ell\varphi + \frac{\pi}{2}\left\{1 - \text{sgn}\left[L_p^{|\ell|}\left(\frac{2r^2}{w_0^2}\right)\right]\right\} \tag{7}$$

where $L_p^{|\ell|}$ is associated Laguerre polynomial and $w_0$ the beam waist of the generated LG beam.

In our case, the SPP plays the fundamental key-role to generate the light beam decoding the specifically designed hologram, expanding the range of possible application whenever information needs to be stored with increased security and counterfeit prevention. Samples have been fabricated by electron-beam lithography on polymethylmethacrylate (PMMA) resist layer, spun over a glass substrate, in high-resolution mode, providing high-quality phase-only diffractive optics. In addition, since this fabrication technique, however extremely precise, is slow and expensive, we investigated the possibility to replicate the fabricated optics with faster mass-production techniques, such as UV lithography [24], which allows higher throughput and much lower production costs. The optical response has been tested on an optical table, showing a correct reconstruction of the encoded information under illumination with the expected OAM field. Conversely, if the hologram is illuminated with a common Gaussian beam, the noise is too high to make the image recognizable.

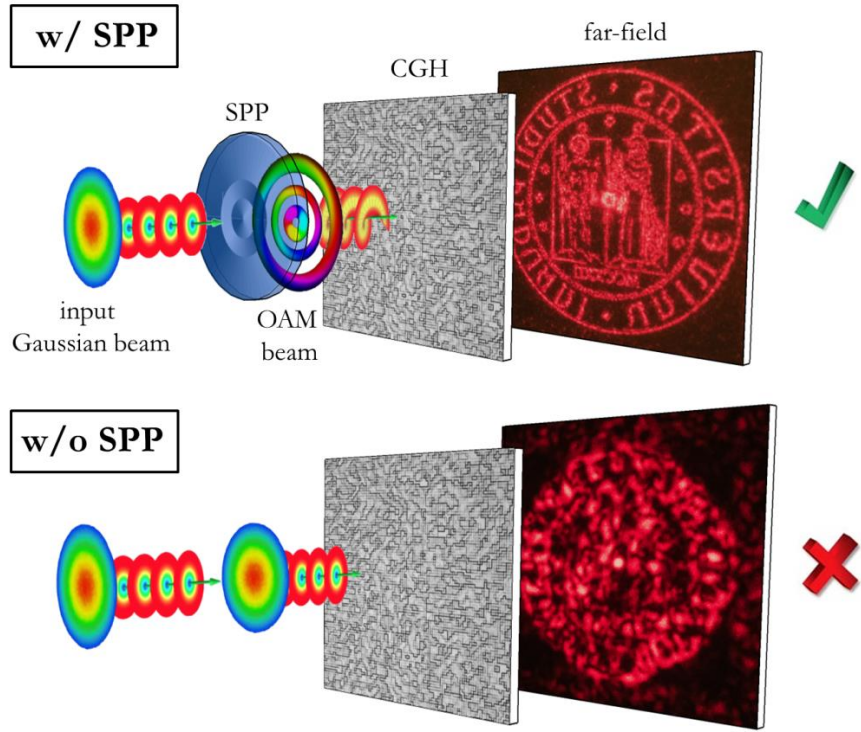

**Figure 2.** Computer-generated holograms (CGH) working principle. Decoding of the CGH with the correct structured light generated by the SPP: the encoded image appears in far-field (at the top). Illumination with standard Gaussian illumination: the image in far field is not recognizable (at the bottom).

## RESULTS

*Holographic design and computation*

The realization of a computer-generated hologram can be schematically split into three steps: analysis, implementation and fabrication. The first one consists in the understanding of the physical process governing the formation of the image encoded on the holographic substrate. Therefore, the hologram phase pattern is implemented through the development of a numerical algorithm, which considers both the physics governing the image formation and the restrictions imposed by the selected fabrication process, e.g. limited resolution and spatial mesh. Finally, the designed pattern is realized, with the properly selected fabrication techniques and protocols.

An Iterative Fourier Transform Algorithm (IFTA) represents a proper choice for the design of computer-generated holograms, due to the capability of generating an optimized phase pattern by bouncing back and forth the information between two spaces related by a Fourier transform, i.e. the hologram plane and the image plane. In Figure 3(d), a scheme of the IFTA algorithm implemented in MATLAB® environment, is shown.

The process begins by collecting the amplitude field generated by the selected SPP. The collected intensity distribution and the corresponding azimuthal phase gradient define the complex input field for the computation of the hologram phase pattern.

We chose three different images with increasing complexity for the design of the associated CGHs. The first one exhibits the official Logo of the University of Padua and is characterized by having a bitmap format with pure black and white pixels (Figure 3(a)). The second one shows two intersecting 'H' characterized by having grayscale 8bit/channel (Figure 3(b)), while the third one represents a wolf portrait with a 8 bit/channel grayscale showing many finer details (Figure 3(c)). In all cases, the input image is centered in the signal window with a size of 200x200 pixels, where the hologram total size is composed of 400x400 pixels.

Starting from the input signal enclosed by a signal window on the image plane, the far-field is brought back to the hologram plane using the inverse Fast Fourier Transform and normalized with respect to the incident illumination, as suggested by eq. (2). The quadratic term in eq. (3) is included in the azimuthal phase of the input field, therefore the image of the computed Fresnel hologram will be at focus on a plane at a distance $z$ from the hologram (fixed at 40 cm).

Within this iteration procedure, the numerical algorithm generates a continuous complex spectrum in the holographic plane. Since the selected lithographic protocol can reproduce phase-only patterns with discretized values of phase, this restriction must be considered and properly implemented in the code. Therefore, a quantization of the phase values is introduced in the IFTA process, within each iteration step, through a direct amplitude-elimination and a direct partial quantization of phase on the hologram plane [25] (see Supplementary Information S2 for more details).

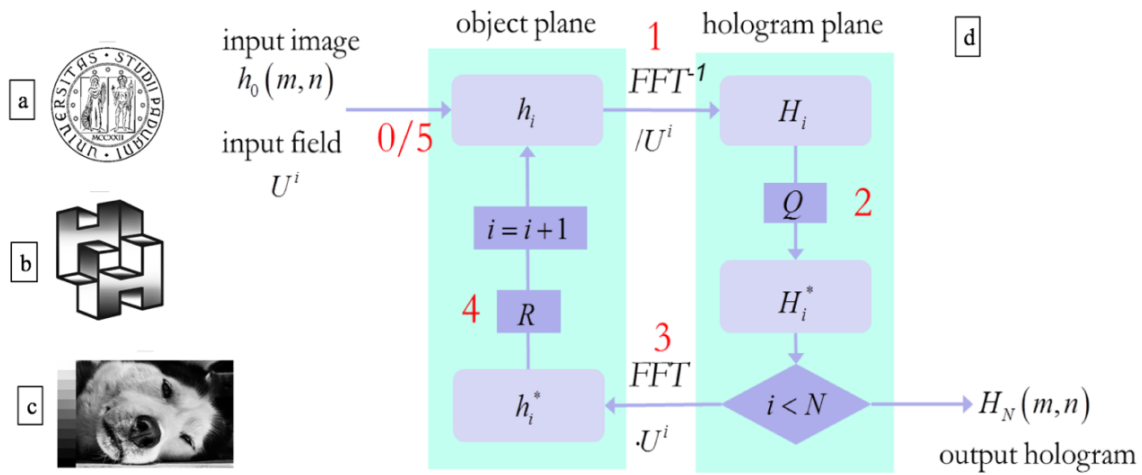

**Figure 3.** Processed images for the Computer-generated holograms. (a) Unipd Logo bitmap format with pure black and white pixel, (b) two intersecting 'H' with a grayscale 8bit/channel and (c) wolf portrait characterized by a 8 bit/channel grayscale with much more fine details. (d) Schematic representation of the iterative Fourier transform algorithm. After the signal input of the *i*th iteration step in the object domain enters the loop, the inverse fast Fourier transform (FFT$^{-1}$) allows the transition to the hologram plane with its transform $H_i$ (1), normalized with respect to the incident field $U^i$, hence the quantization operator $Q$ allows both direct partial quantization of phase and amplitude elimination (2). The discretized hologram pattern $H_i^*$ is obtained and the fast Fourier transform (FFT) is performed (3). From the simulated reconstruction in the object domain, the final new signal $h_{i+1}$ is obtained with the proper replacement of the output signal amplitude with the desired image amplitude within the signal window (4). The loop is repeated (5) for $N$ iterations, until convergence.

This quantization process is obviously expected to affect the final quality of the image, introducing noise in the reconstruction plane. However, this noise can be reduced by replacing the amplitude within the signal window with the desired amplitude of the original signal, while leaving both phase and amplitude free outside the signal window, where the noise is substantially relegated. Then, the loop is repeated using the output signal as the input field for the next iteration step (see figure 4).

With the progress of the iterations, the algorithm converges to an optimized design of the holographic pattern. The convergence can be checked in real-time evaluating the signal to noise ratio (SNR) and the diffraction efficiency ($\eta$) [26]. The SNR gives the correlation between input signal and reconstructed signal and is defined as:

$$SNR_i = \frac{\sum_{m,n} |h_{i,mn}|^2}{\sum_{m,n} |h_{i,mn} - h_{0,mn}|^2} \qquad (8)$$

where $h_{i,mn}$ is the signal at the pixel ($m$, $n$) at the $i$th step and $h_{0,mn}$ is the corresponding signal in the reference image. The diffraction efficiency $\eta$ is typically defined as the ratio of the signal energy within the signal window chosen as an active part of the reconstruction plane, and the total energy in the reconstruction (object) plane, given by:

$$\eta_i = \frac{\sum_{m,n \in SW} |h_{i,mn}|^2}{\sum_{m,n} |h_{i,mn}|^2} \qquad (9)$$

where $W$ stands for the set of pixels inside the signal window.

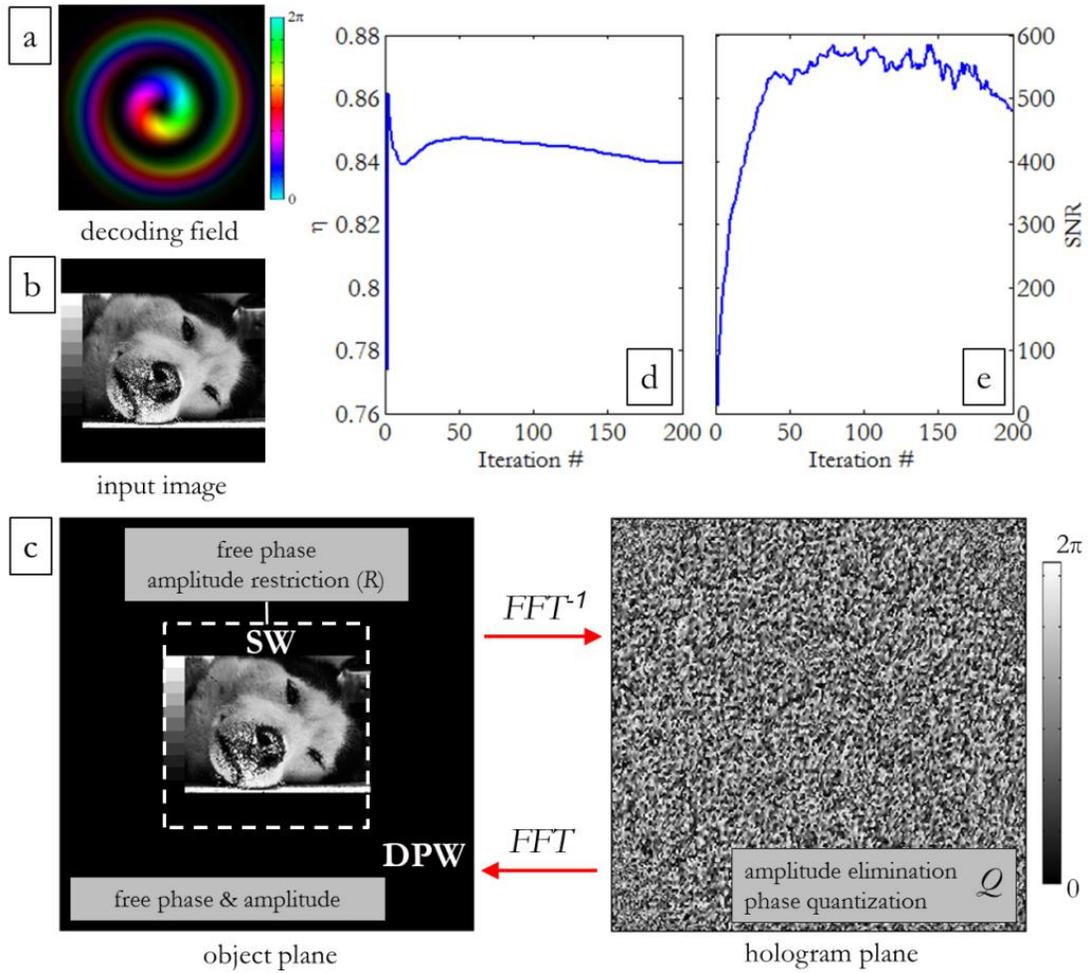

**Figure 4.** (a) Experimental beam generated with SPP with $p=1$, $\ell=1$. Brightness and colors refer to intensity (experimental) and phase respectively. (b) Input image entering the optimization loop. (c) On the left: diffracted pattern window (DPW) on the object plane, including the signal window (SW) where the image is formed. At each step of the algorithm, a replacement of the output signal amplitude with the desired image amplitude within the SW is imposed, while free phase is left on the whole DPW. On the right: corresponding hologram phase pattern after 200 iterations. On the hologram plane, direct amplitude-elimination and partial phase-quantization is performed at each step. Evolution of diffraction efficiency $\eta$ (d) and signal-to-noise ratio (SNR) (e) during algorithm convergence to an optimized design of the computer-generated hologram for illumination with an OAM beam ($p=1$, $\ell=1$) (c).

*Fabrication*

Phase-only diffractive optical elements are fabricated as surface-relief patterns of pixels. This 3-D structures can be realized by shaping a layer of transparent material, imposing a direct proportionality between the thickness of the material and the local phase delay. Electron beam lithography is the ideal technique to fabricate 3D high resolution profiles [27-28]. By modulating the local dose distribution, a different dissolution rate is induced in the exposed polymer, giving rise to different resist thicknesses after the development process. In this work, the SPP and DOE patterns were written on a PMMA resist layer with a JBX-6300FS JEOL EBL machine, 5 nm resolution, working at 100 KeV with a current of 100 pA. The substrate used for fabrication is glass-coated ITO with low surface resistivity (8-12 Ω) in order to ensure a good discharge of the sample during electron beam lithography. After the exposure, the resist is developed in a temperature-controlled developer bath for 60 s.

At the experimental wavelength of the laser ($\lambda$=632.8 nm), PMMA refractive index results $n_{PMMA}$=1.489 from spectroscopic ellipsometry analysis (J.A. Woollam VASE, 0.3 nm spectral resolution, 0.005° angular resolution). The height $d_k$ of the pixels of the $k$th layer for normal incidence in air is given by

$$d_k = \frac{k-1}{M} \frac{\lambda}{n_{PMMA}-1} \tag{10}$$

being $M$ the total number of phase levels, $k = 1...,M$. The fabricated CGHs are 400 x 400 pixels square matrices with $M$=16 phase levels. Each pixel is 3.125 x 3.125 µm², therefore the total area of each sample is 1.250 x 1.250 mm².

Inserting the given laser wavelength and PMMA refractive index in the previous equation, we get: $d_1$=0 nm, $d_{16}$=1213.19 nm, $\Delta d$=80.88 nm. The quality of the fabricated structures has been assessed using optical microscopy (Figures 5(a)), Scanning Electron Microscopy (SEM) (Figure 5(b)) and Atomic

Force Microscopy (AFM) (Figure 5(c-d)). Experimental height values have been compared with the nominal ones exhibiting a remarkable accordance within the experimental errors, estimated by considering surface roughness (see Supplementary Figure 1). Spiral phase plates have been fabricated in PMMA on a transparent glass substrate with the same lithographic process, defining the spiral phase ramp with 256 levels [23]. The total thickness, regarding to the given wavelength and PMMA refractive index, is 1294.1 nm.

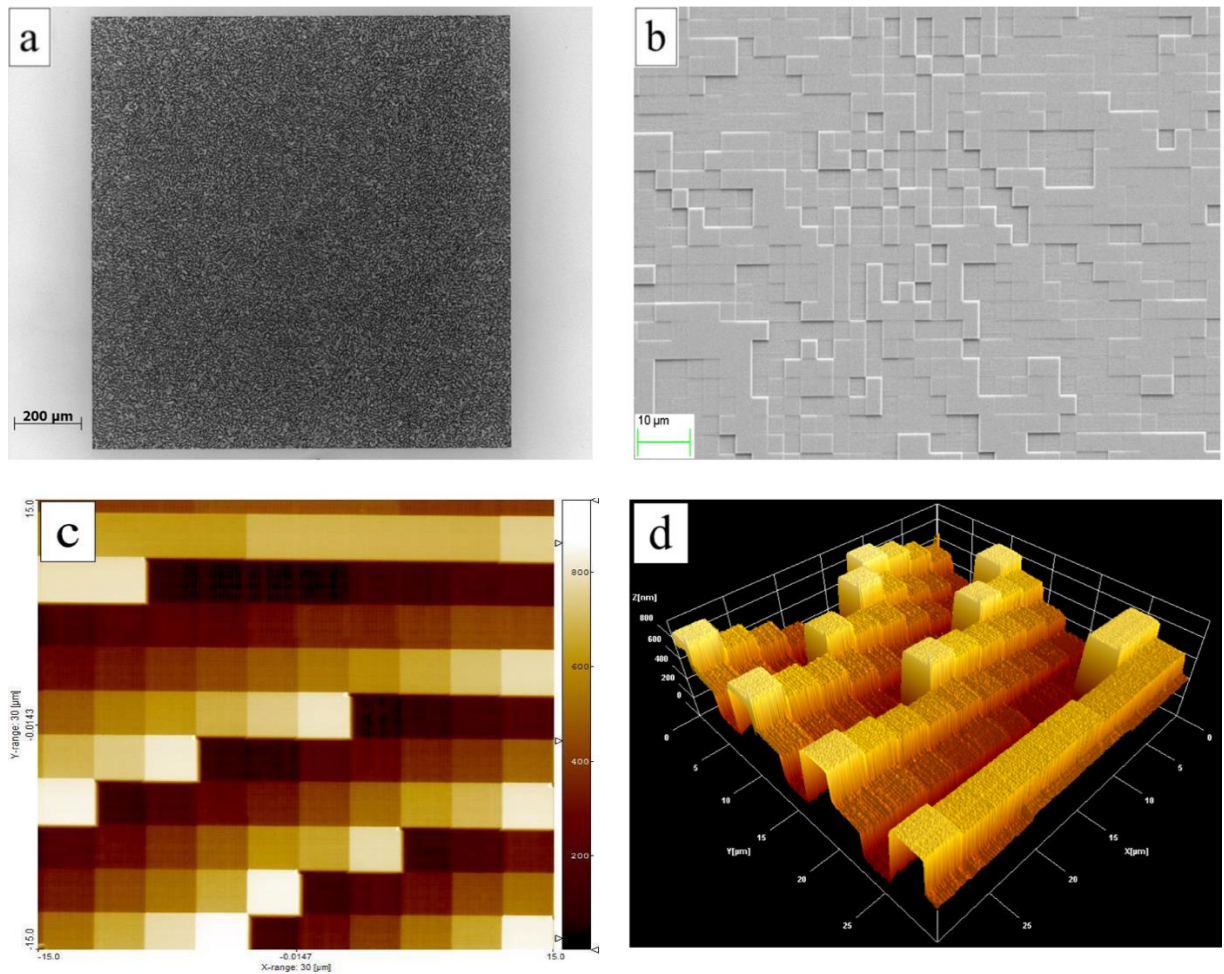

**Figure 5**. Optical microscopy (a), scanning electron microscopy (b) and atomic force microscopy (c-d) of a PMMA phase-only CGH for pixel size: 3.125 x 3.125 μm$^2$. CGH total size: 1.250 x 1.250 mm$^2$. Working wavelength $\lambda$ = 632.8 nm. 16 phase levels.

*Optical characterization*

The optical characterization setup was mounted on an optical table (see scheme in Figure 6). The Gaussian beam was emitted by a HeNe laser source (HNR008R, Thorlabs, $\lambda=$ 632.8 nm, waist $w_0$=240 µm, power 0.8 mW). The polarized beam (LPVISE100-A, Thorlabs) was resized and focused on the selected spiral phase plate. By adjusting the distances from the laser source to the first lens of focal length $f_1$=25 cm and the SPP, the beam-waist is reduced to $w_1$=130 µm. Then the transmitted OAM beam was collimated by a second lens of focal length $f_2$=7.5 cm and a beam-splitter was used to both collect the intensity profile of the generated OAM beam content and to correctly illuminate the holographic pattern. The field profile was collected with a CCD camera (DCC1545M, Thorlabs, 1280 x 1024 pixels, 5.2 µm pixel size, monochrome, 8-bit depth). Far-field images were collected using Nikon D750 camera.

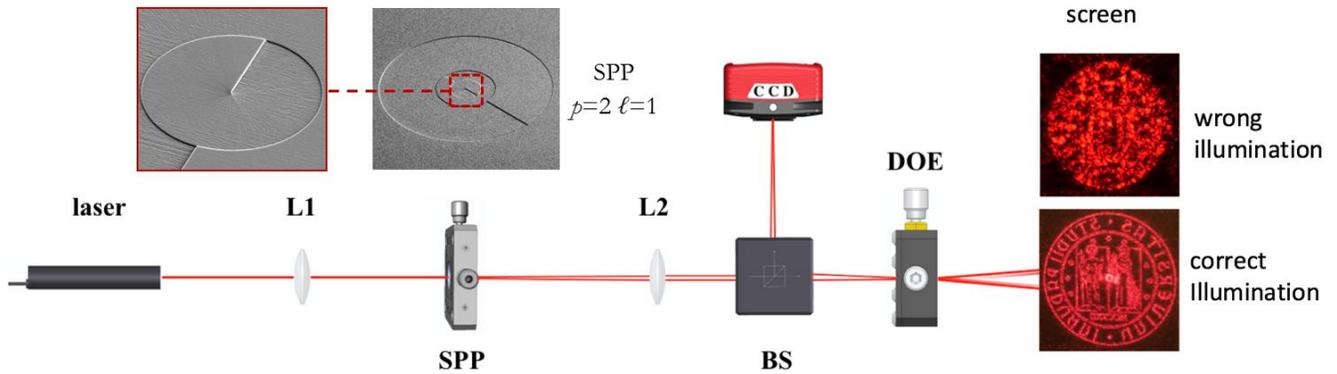

**Figure 6**. Scheme of the optical characterization setup. Laser source ($\lambda$=632.8 nm), first lens (L1), spiral phase plate (SPP), second lens (L2), beam splitter (BS), phase only diffractive optical element (DOE), camera for analysis of the DOE input beam (CCD), screen for revealing the decoded information. Experimental images are reported for a hologram encoding Unipd logo, in case of correct ($p$=1, $\ell$=1) and wrong Gaussian illumination. Inset SEM images: details of SPP for the generation of OAM beam with $p$=2, $\ell$=1.

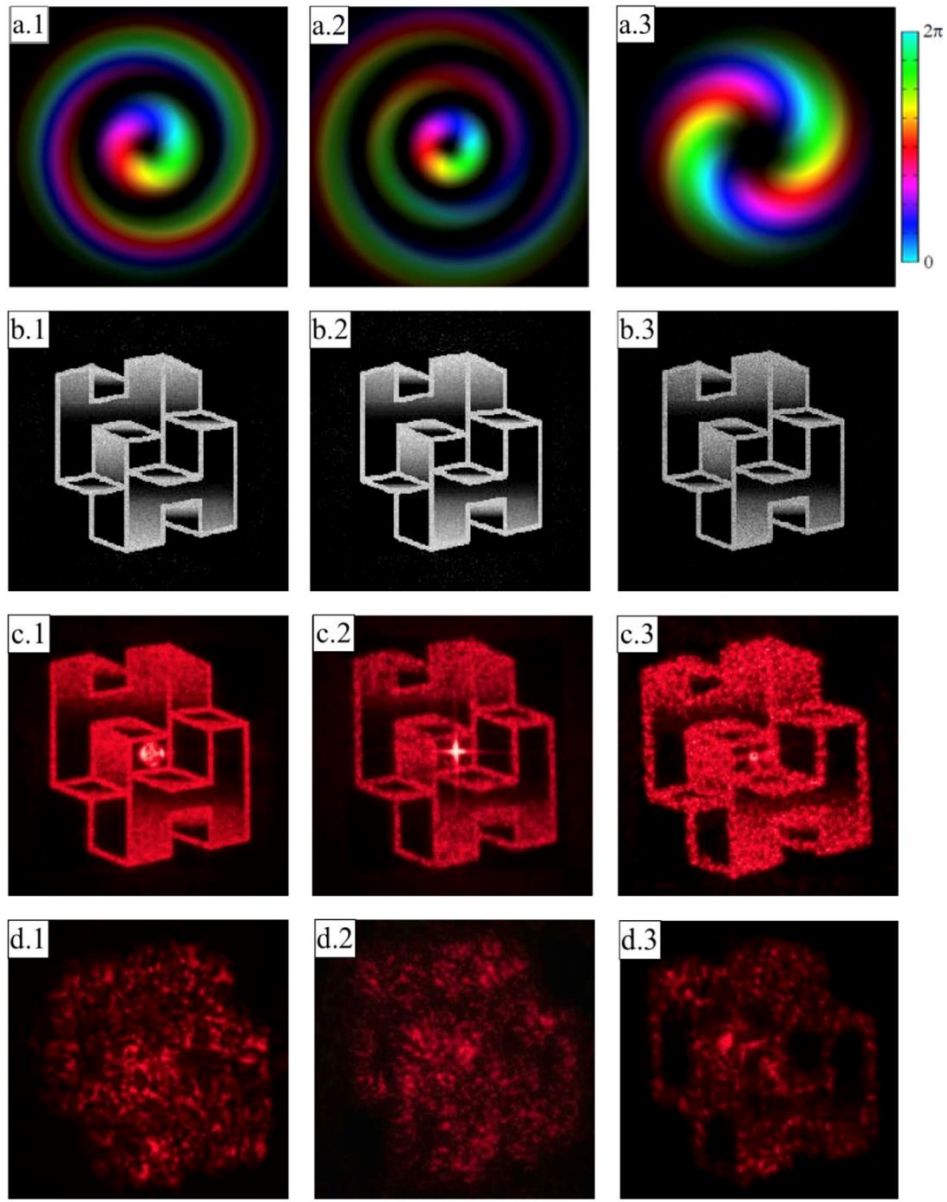

**Figure 7**. (a) Experimental beam generated with SPPs for $p=1$, $\ell=1$ (a.1), $p=2$, $\ell=1$ (a.2), $p=0$, $\ell=2$ (a.3). Brightness and colours refer to intensity (experimental) and phase respectively. (b) Simulated image for correct illumination with $p=1$, $\ell=1$ (b.1), $p=2$, $\ell=1$ (b.2), $p=0$, $\ell=2$ (b.3). (c) Experimental output in case of correct illumination with $p=1$, $\ell=1$ (c.1), $p=2$, $\ell=1$ (c.2), $p=0$, $\ell=2$ (c.3) or wrong Gaussian illumination $p=0$, $\ell=0$ (d.1, d.2, d.3)

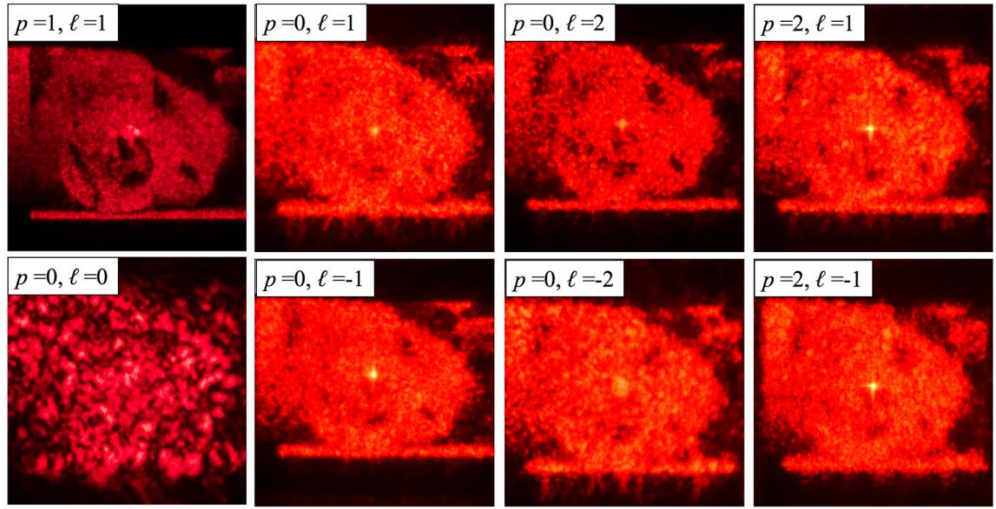

**Figure 8.** Experimental output of the hologram encoding 8 bit/channel grayscale picture, designed for $(p, \ell)=(1, 1)$, under illumination with several spiral phase plates. While the encoded image clearly appears, as expected, when the hologram is illuminated by the correct SPP, under wrong SPP illumination the image is no longer recognizable.

A soft-lithography replica of a holographic PMMA master has been optimized in order to set-up an easy, fast and low-cost fabrication procedure. The optical and morphological characterizations of the generated copies demonstrate exceptional reliability in replicating the hologram 3D structures (see Figure 9 and supplementary information Figure S3).

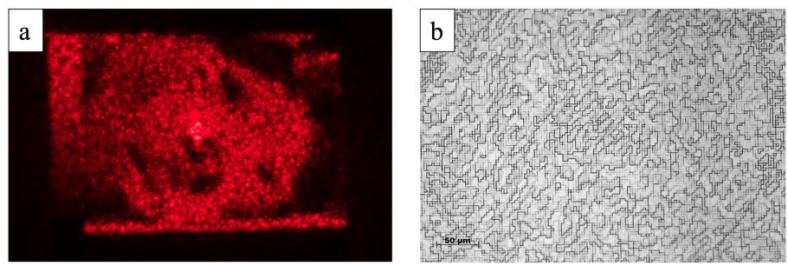

**Figure 9.** (a) Far-field images from holographic-master replica under correct OAM illumination ($p=1$, $\ell=1$), and (b) optical microscope inspection.

In addition, in order to further enhance the possibility of future applications in trace and track using these diffractive optical elements, an image representing a QR-code has been computed and tested under OAM illumination (see supplementary information Figure S2).

DISCUSSION

In this work, we present the first attempt of a complete design and fabrication procedure of computer-generated holograms encoding information for illumination with structured light beams carrying specific distributions of intensity and phase. In particular, we consider beams carrying orbital angular momentum, generated with high-order spiral phase plates enabling both the transfer of topological charge to the input beam and the generation of a multi-ring intensity pattern. An iterative Fourier transform algorithm has been implemented for the computation of an optimized phase pattern for the selected input image and incident field. The computation of the hologram pattern for the given input beam imposes a one-to-one correspondence between the generated hologram and the SPP and therefore increases the security level of these diffractive optics, so that the encoded information cannot be addressed without the correct illumination key. This result has been demonstrated with the design and optical test of several samples, fabricated with high-resolution electron-beam lithography. The optical characterization demonstrates that the encoded image appears, as expected, only when the hologram is illuminated with the correct input illumination, otherwise the information is not recognizable, with neither with standard Gaussian illumination nor different OAM beams. Electron-beam lithography provides a high-resolution lithographic technique, allowing the fabrication of concealed security holograms with a resolution far higher than common dot-matrix optical devices [29, 30], usually fabricated with interferential-lithography systems.

One may consider the cost of the EBL as a drawback, but in fact it is the cost of the increased quality and resolution of the holograms. Moreover, we have demonstrated the replica process of one high-quality hologram to produce several identical copies by soft lithography methods, which are well-considered as a super-economical technique.

In supplementary information S4, the replica process of a holographic PMMA master is explained in details. Soft lithography technique provides an exceptional reliability in reproducing the exact 3D structures in an easy, fast and low cost process.

The experimental optical bench exploited in this work for the optical characterization of the fabricated holograms, is clearly cumbersome and unhandy in the view of commercial and industrial applications. However, further improvements and integration can be performed. In particular, the SPP can be embodied to the laser source exploited for hologram inspection. The SPP phase pattern will be properly optimized for the size of the beam exiting the laser, and a lens term can be integrated in the SPP phase profile in order to focus the OAM beam at a proper distance, without the need of a further lens as shown in Figure 6.

By engineering this new type of diffractive optical elements, which correctly decode visual information only when illuminated with light owning a specific spatial distribution of intensity and phase, we expand the available range of security optical devices and of possible applications of structured light, whenever information needs storing with increased security and counterfeit prevention. The miniaturized size of the fabricated optics allows them to be either integrated or concealed on greater-size optical elements, such as 2D/3D holograms and other types of overt security devices to be applied on documents and products.

METHODS

*Numerical simulations.*

A custom MATLAB code based on IFTA was implemented in order to compute the phase-only diffractive optical element specifically designed for OAM illumination.

*Electron beam lithography.*

All 3D multilevel structures have been fabricated in a 2 μm thick PMMA resist with a molecular weight of 950 k (kg/mol), spin-coated on a 1.1 mm thick ITO coated soda lime float glass substrate and pre-baked for 10 min at 180°C on a hot plate. For the grey-scale lithography step, a dose-depth correlation (contrast curve) was used. Contact profilometry was performed to determine the remaining resist heights. Dose-to-clear value (complete removal of PMMA) was found to be 566 μC/cm$^2$. CGH patterns were written with a JBX-6300FS JEOL EBL machine, 12 MHz, 5 nm resolution, working at 100 KeV with a current of 100 pA. The presence of the ITO layer was necessary in order to ensure a good discharge of the sample during electron beam lithography. A dose correction for the compensation of proximity effects has been applied. This compensation is required both to match layout depth with the fabricated relief and to obtain a good shape definition, especially in correspondence of the phase steps. Exposed samples were developed under slight agitation in a temperature-controlled developer bath for 60 s. Deionized water: isopropyl alcohol (IPA) 3:7 was found to be the most suitable developer, giving optimized sensitivity and contrast characteristics as well as a minimized pattern surface roughness at 20 °C. After development, the samples were gently rinsed in deionized water and blow-dried using nitrogen flux. Different techniques have been used in order to assess sample quality: tapping-mode atomic force microscopy, optical microscopy and scanning electron microscopy.

*Soft-lithography replica*

A suitable amount of the elastomer Sylgard 184 polydimethylsiloxane (PDMS) base is mixed with the catalyst in a weight ratio of 10:1 respectively, stirred for a while, and then cast onto the surface of the EBL fabricated master. During the PDMS pouring and mixing, creation of bubbles is inevitable, so the container is placed in a desiccator for 30-45 minutes to de-gas it and remove the trapped air bubbles. The sunken master with PDMS prepolymer is then placed in an oven at the temperature of 100°C to be cured for 35 minutes, then was left in a freezer for 5 to 10 minutes to cool down. This shrinks the PDMS slightly and helps when peeling the samples out of their molds.

To fabricate the replica of the mold, a UV-curable photopolymer (Norland Optical Adhesive 74) is dropped onto a glass substrate. The PDMS mold is overlaid on it (with the patterned side facing the liquid) and is pushed a bit to form the contact. Finally, the photopolymer is cured under UV light for 30 seconds, after which the PDMS mold is carefully peeled off to get the final replica.

*Optical characterization*

The characterization setup was designed and assembled on an optical table with gimbal piston isolators (refer to Figure 6). The Gaussian beam was emitted by a HeNe laser source (HNR008R, Thorlabs, $\lambda=$ 632.8 nm, waist $w_0$=240 µm, power 0.8 mW). The polarized beam (LPVISE100-A, Thorlabs) impinges on the corresponding spiral phase plates, mounted on a sample holder with micrometric drives (ST1XY-S/M, Thorlabs, travel 2.5 mm, resolution 10 µm). By adjusting the distances from the laser source to the first lens of focal length $f_1$=25 cm and the SPP, the beam-waist was reduced to $w_1$=130 µm. Then the transmitted beam was collimated by a second lens of focal length $f_2$=75 cm and a 50:50 beam-splitter was used to both collect the intensity profile of the generated OAM beam content and to correctly illuminate the holographic pattern. The field profile was collected with a CCD camera (DCC1545M, Thorlabs, 1280 x 1024 pixels, 5.2 μm pixel size, monochrome, 8-bit depth). The holographic sample was fixed on a vertical *XY* translation mount with micrometric drives (ST1XY-S/M, Thorlabs, travel 2.5 mm, resolution 10 μm), and the far-field was projected on a white screen. The projected far-field images were collected using Nikon D750 camera.

ACKNOWLEDGEMENTS

This study has been supported by FSE project HOLOAM 2105-84-2121-2015 funded by Veneto Region. The authors gratefully thank Maurizio Motta of Holo3D Srl for the interesting discussions during this work.

AUTHOR CONTRIBUTION STATEMENT

G.R. developed the numerical codes for holograms phase-pattern calculation and design for the lithographic process. R.R. performed holograms design and optical characterization and test. M.M. carried out the fabrication with electron-beam lithography and performed microscopy and AFM characterizations. E.M. realized soft-lithography replica. P.C. performed AFM and SEM analyses. F.R. proposed and supervised the project.

# Supplementary information

S1. AFM ANALYSIS

Atomic Force Atomic (AFM) microscopy has been performed in tapping-mode configuration for a 16-level phase-only computer-generated hologram (CGH) in PMMA, working in transmission at $\lambda=632.8$ nm. Experimental depth values are: $d_0=0$ nm, $d_1=82.7$ nm, $d_2=161.2$ nm, $d_3=248.8$ nm, $d_4=325.9$ nm, $d_5=401.0$ nm, $d_6=494.0$ nm, $d_7=571.0$ nm, $d_8=636.0$ nm, $d_9=721.0$ nm, $d_{10}=810.0$ nm, $d_{11}=885.0$ nm, $d_{12}=973.0$ nm, $d_{13}=1055.0$, $d_{14}=1138.0$ nm, $d_{15}=1219.0$ nm, $d_{16}=1291.0$ nm. In Supplementary figure 1, a 3D AFM reconstruction is reported for a small area of 30 x 30 µm².

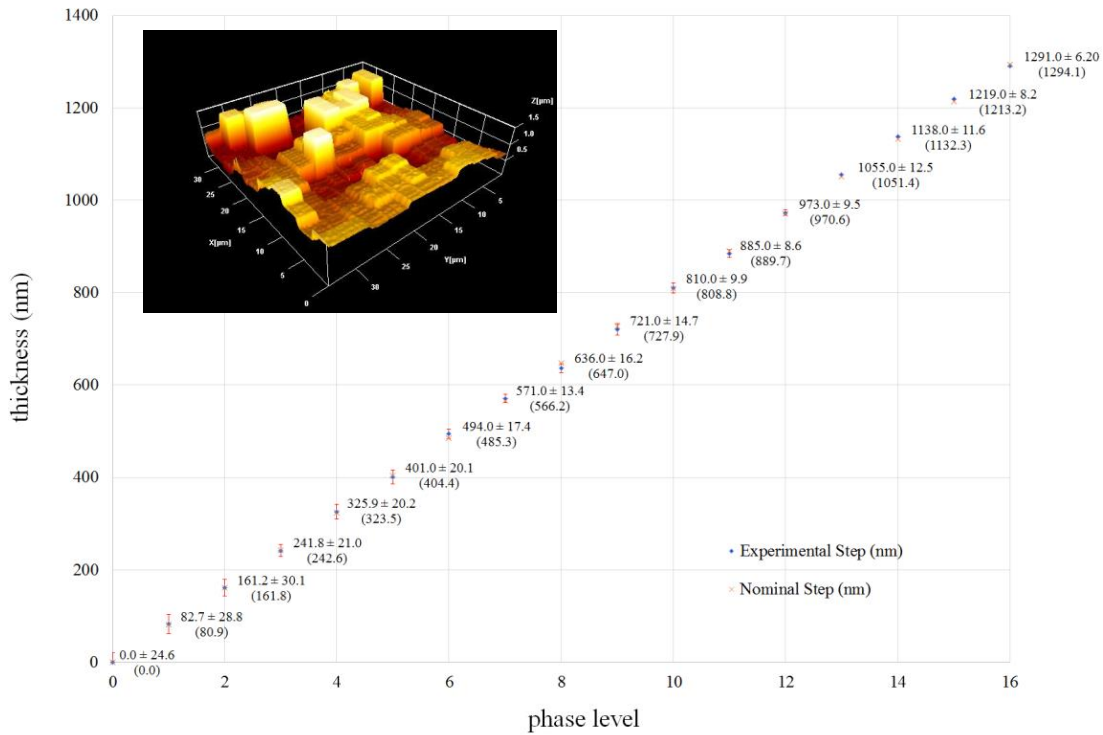

**Supplementary Figure 1**. Analysis of 30 x 30 µm² (spanning about 10 x 10 pixels) zone of the 16-level holographic DOE. Comparison between experimental (rhombus with error bars) and nominal (cross) heights is reported.

Experimental values are compared with the nominal ones exhibiting a remarkable accordance within the experimental errors, estimated by considering surface roughness. Roughness root-mean-square (RMS) increases from 6 nm to 25 nm, for the lowest and highest depth level respectively.

S2. ITERATIVE FOURIER TRANSFORM ALGORITHM (IFTA)

The main task of the iterative algorithm is to transform the complex hologram spectrum function into a discrete form (to a binary or $2^n$-level phase-only form). This discretization is done sequentially during the process. Obviously such quantization introduces noise in the reconstruction plane, which is removed, or at least significantly reduced, by iteratively replacing the amplitude of a newly calculated noisy signal within the signal window with the desired amplitude of the original signal. This modified signal is used as input for the following step. The quantization represents a transformation of the hologram function to the desired form while maintaining reconstructed image quality within defined constraints. In mathematical terms, the whole IFTA procedure can be resumed as it follows:

1. $h_i$ is the signal at the input of the $i$th iteration loop in the object plane. After normalizing with respect to the incident field $U^i$, comprehensive of the Fresnel term, the inverse fast Fourier transform allows passing to the hologram plane with its transform $H_i$:

$$H_i = FFT^{-1}\left[h_i / U^i\right] \quad (1)$$

2. The quantization operator $Q$ is applied in the hologram space and the phase function $H_i^*$ is generated:

$$H_i^* = Q[H_i] \quad (2)$$

When iteratively shaping multilevel profiles within the IFTA method, there remains an open questions of how to eliminate iteratively the amplitude of the spectrum and how to correctly perform phase quantization. The operator $Q$ should, in principle, consist of two parts:

$$Q = P \cdot A \tag{3}$$

where $P$ and $A$ are respectively the phase and the amplitude operators, which performs the quantization at each iteration of the hologram phase and amplitude respectively. As far as the amplitude operator is concerned, the *direct (single-step) amplitude elimination* is usually applied, with very good results, that is:

$$A[x] = \frac{x}{|x|} \tag{4}$$

A correct choice of $P$ operator is crucial for convergence of the algorithm and optimization of the hologram. Phase quantization is performed by iterative expansion of local surroundings of quantization levels, and is represented formally by the operator $P$. This operator has the form:

$$P[x] = \begin{cases} q_j & x \in \varepsilon_j(i) \\ x & otherwise \end{cases} \tag{5}$$

where $\{q_j\}$ is the set of $M$ target quantization points lying on the unitary circle in the complex plane and $\{\varepsilon_j(i)\}$ is the set of particular surroundings within the $i$th iteration cycle that simultaneously defines the areas where the replacement of the values takes place. Each surrounding is defined as:

$$\varepsilon_j(i) = \{q \mid |angle(q) - angle(q_j)| \leq e_j(i)/2\} \tag{6}$$

where $e_j(i)$ is the amplitude of the surrounding $\varepsilon_j(i)$, centered on the quantization phase level $angle(q_j) = atan(Im(q_j)/Re(q_j))$, at the $i$th step. The idea is, at each iteration, to leave a certain number of unprocessed values outside regions adjacent to the quantization levels. In order to make further optimization feasible, it is necessary to keep a sufficient number of these unprocessed values during the process and enable values already processed to jump from one target phase level to either another phase level or to the regions of unprocessed values. On the one hand the quantization ranges $e_j(i)$ can be kept fixed throughout the iteration, equal to the whole spectral range of interest for the given set of quantization levels. In this case a full quantization only would be periodically repeated. On the other hand, $e_j(i)$ can be enlarged gradually step by step, while each actual size of the surroundings is applied only once, that is:

$$e_j(i) = \frac{2\pi}{M} \frac{i}{N} \qquad (7)$$

This procedure is known as *one-step approach with partial phase quantization*. Alternatively, this procedure can be iterated, in the form of the two-step approach or repeated whole partial quantization. A further scheme consists in the repetition of each partial quantization step: two-step approach with repetition of each partially quantizing step. In this last case all expanded surroundings are themselves repeated while the whole quantization is performed once. In this work we applied direct amplitude elimination and one-step partial phase quantization.

3. The simulated reconstruction in the object domain $h_i^*$ is given by the fast Fourier transform, once the hologram has been multiplied by the incident field:

$$h_i^* = FFT\left[H_i^* \cdot U^i\right] \qquad (8)$$

4. Finally the new signal $h_{i+1}$ is obtained by the proper replacement of the output signal $h_i^*$ with the desired signal within the object window, through the operator $R$:

$$h_{i+1} = R\left[h_i^*\right] \qquad (9)$$

At this step, the aim is to disturb the partially quantized structure in the smallest possible way. This signal modification, in fact, partially destroys the quantized character of the structure already formed. This detrimental effect should be minimized using a scale factor $\gamma$:

$$\gamma_i = \frac{\sum_{m,n} FFT\left[w \cdot h_i^*\right] FFT^\dagger \left[w \cdot h_0 \cdot \frac{h_i^*}{|h_i^*|}\right]}{\sum_{m,n \in W} |h_0|^2} \tag{10}$$

where $w=1$ inside the signal window, otherwise it is null.

5. The current loop is completed and the iteration goes on until the total number $N$ of iteration is reached

## S3. OPTICAL CHARACTERIZATION OF A QR-CODE AS HOLOGRAPHIC IMAGE.

In order to further enhance the range of applications in anti-counterfeiting and brand protection, a holographic QR-code has been used as image and characterized on the optical bench (refer to Figure 6 for the optical setup).

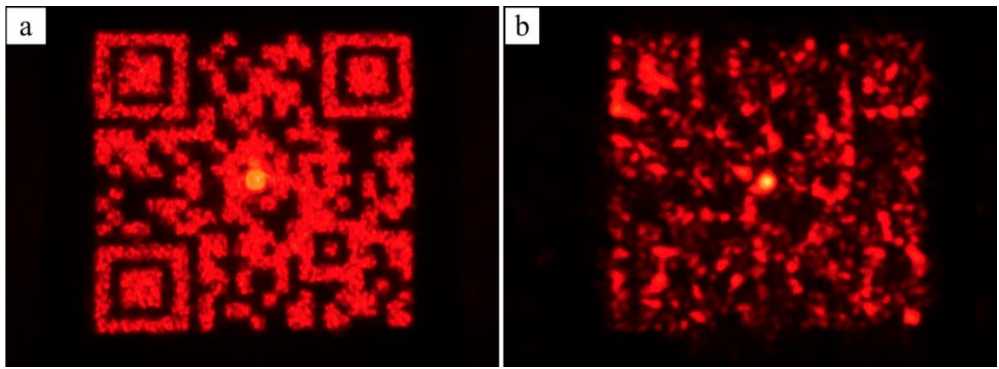

**Supplementary Figure 2.** Experimental output of a QR-code holographic image. Case of (a) correct OAM illumination for ($p=1$, $\ell=+1$) or wrong (b) Gaussian illumination ($p=0$, $\ell=0$).

When the image is correctly illuminated, it allows the recognition of the http:// address using common mobile APPs such as (QR code reader, Android App). In the presented case, the associated address corresponds to http://www.lann.it

## S4. REPLICA WITH SOFT LITHOGRAPHY OF HOLOGRAPHIC DOEs

Fabrication techniques like photolithography, electron-beam lithography, etc. are extremely successful in patterning structures of radiation-sensitive materials (e.g. photoresists or electron-beam resists) on glass or semiconductor surfaces. However significant challenges exist in adapting these lithographic methods for emerging applications and areas of research that require unusual systems and materials (e.g. those in biotechnology, plastic electronics, etc.), large patterned areas or nonplanar surfaces. These established techniques also involve high capital and operational costs. As a result, older and conceptually simpler forms of lithography – embossing, molding, stamping, writing, etc. – are being re-examined for their potential to serve as the basis for nanofabrication techniques that avoid these limitations. These are altogether known as 'soft lithography', named for their use of soft, elastomeric elements in pattern formation.

The soft lithography process can be arranged in two steps: fabrication of the elastomeric elements and use of these elements to pattern features in geometries defined by the element's relief structure. The original patterned structure from which the stamp is derived is known as the 'master'. Many elastomeric elements can be generated from a single master, and each element can be used several times to replicate the initial master. Elastomeric elements are generated by casting a light- or heat-curable prepolymer against the master. With optimized materials and chemistries, this fabrication sequence has remarkably high fidelity. In fact, recent works shows that relief with nanometer depths and widthscan can be reproduced accurately.

In this work, a soft lithography process has been finely tuned in order to realize high-fidelity replicas of computer-generated holograms fabricated by high-resolution electron beam lithography. Particular attention has been paid to the optimization of the curing processes in order to best preserve the steepness of the vertical profiles of the original patterns. In addition, materials have been carefully selected in order to ensure the same optical behaviour of the original masters. The quality of the replicas has been tested with optical and AFM microscopy and characterized with an optical setup mounted on an optical table. The optical behaviour of the fabricated replicas is comparable to that of the original CGH masters. The successful transfer of this technique into manufacturing settings for important applications that cannot be addressed effectively with other lithographic methods offers the promising possibility to realize the holographic optical elements through high-throughput and low-cost realization processes. In Supplementary Figure 3 a scheme of the replica process exploited in this work is reported.

*Fabrication of the elastomeric element* (*PDMS mold*)*:* a suitable amount of the elastomer Sylgard 184 polydimethylsiloxane (PDMS) base is mixed with the catalyst in a weight ratio of 10:1 respectively, stirred for a while, and then cast on the surface of the EBL fabricated master (PMMA mold). During the PDMS pouring and mixing, creation of bubbles is inevitable, so the container is placed in a desiccator for 30-45 minutes to de-gas it and remove the trapped air bubbles. The sunken master with PDMS prepolymer is then placed in an oven at the temperature of 100°C to be cured for 35 minutes, then left in a freezer for 5 to 10 minutes to cool down. This shrinks the PDMS slightly and helps when peeling the samples out of their molds.

*Replica process* (*NOA cast*)*:* to fabricate the replica of the mold, a UV-curable photopolymer (Norland Optical Adhesive 74) is dropped on a glass substrate. The PDMS mold is overlaid on it (with the patterned side facing the liquid) and is pushed a bit to form the contact. Finally, the photopolymer is cured under UV light for 30 seconds, after which the PDMS mold is carefully peeled off to get the final replica.

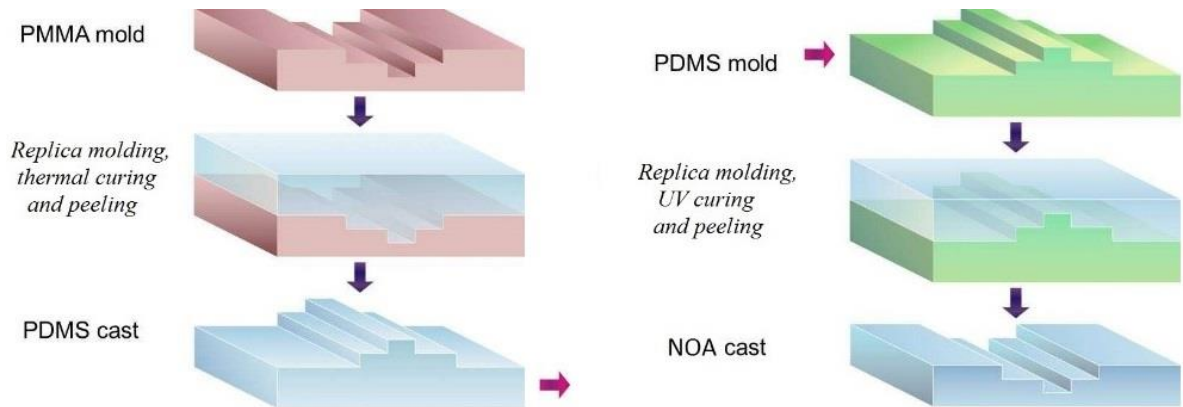

**Supplementary Figure 3.** Essential scheme of the soft lithography process. On the left: PDMS replica of the master. PDMS is cast on the surface of the master, thermal cured and peeled off. On the right: NOA replica of the PDMS mold. The NOA photopolymer is dropped onto the PDMS mold, UV-light cured and peeled off the mold to get the final cast.